\documentstyle[12pt,epsf]{article}
\newcommand{\CA}{{\cal A}}

\newcommand{\CT}{{\cal T}}
\begin{document}
\def\bp{{\mbox{\bf p}}}
\def\bP{{\mbox{\bf P}}}
\def\bD{{\mbox{\bf D}}}
\def\bk{{\mbox{\bf k}}}
\def\br{{\mbox{\bf r}}}
\def\bq{{\mbox{\bf q}}}
\def\bn{{\mbox{\bf n}}}
\def\ba{{\mbox{\bf a}}}
\def\bb{{\mbox{\bf b}}}
\def\bc{{\mbox{\bf c}}}
\def\bxi{{\mbox{\boldmath$\xi$}}}
\def\bsigma{{\mbox{\boldmath$\sigma$}}}
\def\1s0{{^1\!S_0^{++}}}
\def\sqsf{{\sqrt{s_f}}}
\thispagestyle{empty}
\begin{center}
{ \Large \bf
Final State Interaction within the
Bethe-Salpeter Approach in Charge Exchange  $pD\to n(pp)$ process.
}\\[10mm]
\noindent
S.S. Semikh$^{1 \dag}$, L.P. Kaptari$^{1}$, S.M. Dorkin$^{2}$,  and B.
K\"ampfer$^{3}$

\vskip 5mm

{\small
(1) {\it
BLTP, Joint Institute for Nuclear Research, Dubna, Russia
}
\\
(2) {\it
SNPI, MSU, Dubna, Russia
}
\\
(3) {\it
Forschungszentrum Rossendorf, Germany
}
\\
$\dag$ {\it
E-mail: semikh@thsun1.jinr.ru
}}
 \end{center}

\begin{abstract}
The exclusive charge exchange reaction  $pD\to n(pp)$
at intermediate and high energies is studied
within the Bethe-Salpeter formalism.
The final state interaction in the detected
at zero-near excitation energy $pp$-pair
is described by the $^1S_0$ component of the Bethe-Salpeter
amplitude. Results of numerical calculations of polarization
observables and differential cross-section persuade that,
as in the non relativistic case, this reaction can be
utilized for a relativistic deuteron tensor polarimeter
and as a source of  information about the elementary nucleon-nucleon
charge exchange amplitude.
\end{abstract}
\section{Introduction}
\label{introd}

Nowadays  large programs
of experimental study of processes with polarized particles
are in progress.
The setups with deuteron targets (beams) occupy the leading place
\cite{alexa,kox0,preliminar,cosy_proposal}.
For an investigation of the $NN$ interaction in the deuteron
at short distances the three deuteron form factors,
 magnetic, electric and quadrupole,   should be determined.
 In the elastic   $eD$-scattering with unpolarized particles
 one can measure only two
 quantities, e.g. the magnetic form factor and   the deuteron
function  $A(Q^2)$, which is a kinematical combination
of all three form factors. Even these two quantities provide important information
about the quark physics and dynamics at short distances (see, for instance recent
measurements \cite{alexa} at TJNAF). However, for a full determination
of the deuteron form factors, one
needs measurements with polarized particles. For example, measurements of the
tensor analysing power $T_{20}$ of recoil deuterons in  elastic
 $eD$-scattering allow the determination of the charge form factor
 $G_c$ at high transferred momenta.
 The hadron-deuteron processes can be considered
 as complementary tools in investigation of phenomena at short distances and
  also as a source of unique information unavailable in electromagnetic reactions
 (study of nucleon resonances, checking the non relativistic effective models,
$NN$ potentials etc.).
 Experimental and theoretical investigation of the proton-deuteron
 processes at intermediate and high energies
 has started some decades ago by studying elastic $pD$ scattering~\cite{elsatic},
 exclusive and inclusive break-up~\cite{inclusive,cosy}.
 In $pD$ processes it is possible to completely restore the reaction amplitude
 by measuring a full set of polarization observables
 (see, e.g. \cite{rekalo,ladygin,ourprc}). Hence,
  as in the electromagnetic case one needs to measure different polarizations of
  the recoil deuteron. Since polarization  observables can be studied
  only by an  additional secondary scattering
  of the reaction products (in polarimeter), it is obvious that second process
  must possess a high enough cross section to assure a good  efficiency of
  the polarimeter. In ref.~\cite{wil1}
        Bugg and  Wilkin have proposed as an effective deuteron polarimeter
 the process $p{\vec D}\to (pp)n$ where the final  $pp$-pair is
 detected with extremely low excitation energy (see also ref. \cite{polder}).
Later
investigations \cite{ishida,wil1,moto,wil2,kox}
confirmed the theoretical predictions and the charge exchange
 processes were suggested also for use in investigation
of a number of reactions with deuterons, e.g.
   $pp\to D\pi^+$ \cite{bugg},
 $NN\pi$-systems, inelastic  $({\vec D},{\vec D'})$-reactions off
 heavy nuclei  to study isoscalar transitions
  $\Delta T = 0,\, \Delta S = 1$ \cite{morlet} etc.
The direct consequence of these facts is that
 nowadays the
 interest in investigation of charge exchange processes does not abate.
 In our previous paper \cite{nash_yaf} we investigated the process
 $pD\to n(pp)$  within the impulse approximation.
The goal of the present work is to  consider theoretically the effects
of the final state interaction in this reaction at relativistic
initial energies (COSY, Dubna)
 and to confess whether  in this case the non relativistic predictions~\cite{wil1}
 hold and the reaction can be still regarded as a deuteron polarimeter tool.
 We propose a covariant generalization of the spectator mechanism~\cite{wil1}
  based  on the Bethe-Salpeter formalism and on numerical solution of the
 Bethe-Salpeter  (BS) equation
 with a realistic one-boson exchange kernel~\cite{solution,parametrization}.
Here we focus our attention
 in calculations of the cross section and
 the tensor analysing power $T_{20}$ within the COSY kinematical conditions
 at, as in non relativistic case, zero vector analyzing
 powers of the deuteron.

\section{Kinematics and the invariant amplitude} \label{gl1}
\begin{figure}[h]    
\centerline{\epsfbox{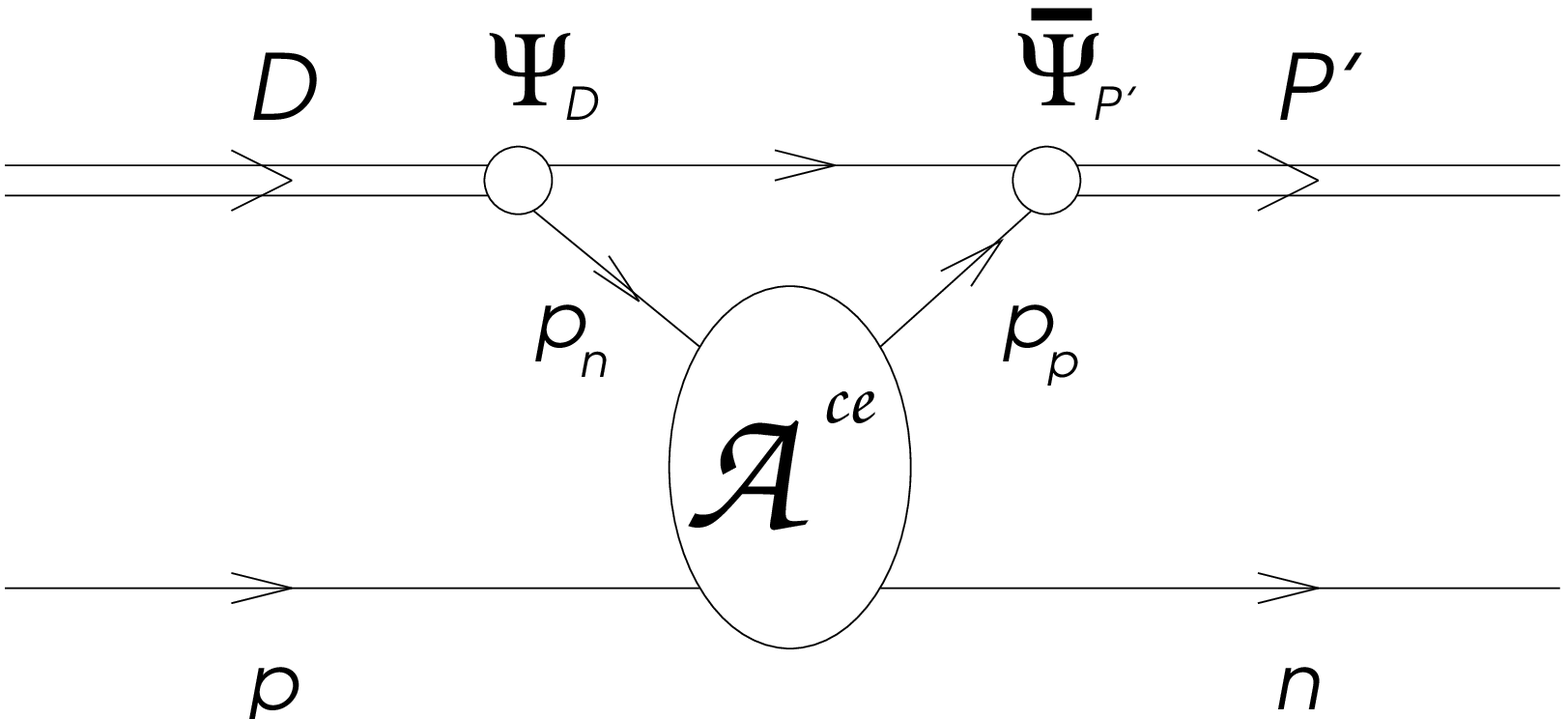}}
\caption{The spectator mechanism for the  charge-exchange process  $Dp\to (pp)n$.
The Bethe-Salpeter amplitudes for the deuteron bound
state and the $pp$-pair in the continuum are denoted as $\Psi$ and $\bar \Psi$,
respectively. The elementary $pn$ charge-exchange amplitude is
symbolically represented by ${\cal A}$.
}
\label{pict1}
\end{figure}

We select  those processes which, in
the deuteron center of mass system,
correspond to final states with one fast neutron and
 a slowly moving
proton-proton  pair, i.e. reactions of the type
\begin{equation} p\,+\,\vec D
\,=\, n + (p_1+p_2).  \label{reaction}
\end{equation}
The transferred momentum from the proton to the neutron
is low, hence the main mechanism of the reaction can be
described as a charge exchange process of the incoming proton off
the internal  neutron whereas the proton in the deuteron remains
merely as a spectator.
If so, then the resulting  $pp$-pair will be detected
with low total and relative momenta.
 In Fig.  \ref{pict1} the diagram
 of such  processes is schematically depicted.
 The following notations are adopted:  $p=(E_p,\bp)$ and
$n=(E_n,\bn)$ are the 4-momenta of the incoming proton and outgoing neutron,
 $P'$ is the total 4-momentum of the  $pp$-pair, which is a sum of the corresponding
 4-momenta of detected protons, $p_1=(E_1,\bp_1)$,
$p_2=(E_2,\bp_2)$: $P'=p_1+p_2$.
The notion of the invariant mass of the pair
 $s_f,~s_f=P'^2=(2m+E_x)^2$, where  $m$ stands for the nucleon
 mass and  $E_x$ for the excitation energy
 of the pair, is also explored in what follows.
 The excitation energy  $E_x$
   ranges from zero to few $MeV$, $E_x\sim 0-8~MeV$. At such
   low values of  $E_x$ the main contribution to the final state
 of the $pp$ pair in the continuum comes from the  the
 $^1S_0$-configuration~\cite{wil2}.
The differential cross section for the reaction (\ref{reaction}) reads
\cite{nash_yaf}:
\begin{equation}
\frac{d^2\sigma}{dt\,ds_f}\,=\,\frac{1}{2}\frac{1}{64\pi\lambda(p,D)}\,
\sqrt{1-\frac{4m^2}{s_f}}\,\frac{1}{(2\pi)^2}\,|M_{fi}|^2.
\label{cross3}
\end{equation}
The experimental data from Saclay \cite{kox}
are binned into the following intervals of
excitation energy:
\begin{eqnarray}
I:   && \quad 0\leq E_x \leq 1~ MeV,    \nonumber\\
II:  && \quad 1~MeV\leq E_x \leq 4~ MeV,\nonumber \\
III: && \quad 4~MeV\leq E_x \leq 8~MeV.\nonumber
\end{eqnarray}
To compare the cross section (\ref{cross3}) with the experimental data
it is necessary to integrate over the invariant mass of the pair in the
regions $I$, $II$ and $III$ according to
\begin{eqnarray}
\label{crint}
\left(\frac{d\sigma}{dt}\right)_{I,II,III}\,=\,
\frac{1}{(8\pi)^3\lambda}\,
\int\limits_{I,II,III}\,ds_p\,\sqrt{1-\frac{4m^2}{s_f}}\,|M_{fi}|^2.
\end{eqnarray}
By using the Mandelstam technique \cite{mandel}
the covariant matrix element corresponding to the diagram on Fig.~\ref{pict1}
 can be written in the form
\begin{eqnarray}
\nonumber
\CT_{r'r}^M &=& \sum\limits_{ss'}\frac{1}{(2m)^2}\int d^4k\,
f_{r's',sr}\times\\
&\times&{\bar u}^{s}(p_n) {\Psi}_M(k)\,(\frac{\hat D}{2}-{\hat k}+m)
{\bar\Psi}_{P'}(k-\frac{q}{2})\,u^{s'}(p_p).
\label{fin}
\end{eqnarray}
The elementary charge-exchange amplitude
$\CA^{ce}$ is incorporated in the matrix element (\ref{fin})
by the on-shell amplitudes $f_{r's',sr}$ and the Dirac spinors.
In doing so
the off-shell effects are neglected and the elementary subprocess is
considered as real process with on-shell particles.
In our numerical calculations we use the
helicity amplitudes of pn scattering,
resulting from both the Nijmegen partial wave analysis
\cite{swart,www} and the well-known results of SAID \cite{said}.
For the final  $^1S_0$-state within the  $\rho$-classification
the BS amplitude ${\bar\Psi}_{P'}$ in the center of mass of the $NN$ pair
is represented by four partial amplitudes
$^1S_0^{++}$, $^1S_0^{--}$, $^3P_0^{+-}$ and $^3P_0^{-+}$ \cite{nashi},
which for the sake  of brevity are denoted as $\phi_1,\dots,\phi_4$.
The partial amplitudes  $\phi_i$ may be found from
the BS equation, which, in the simplest case of
pseudo scalar exchanges reads as
\begin{equation}
\label{neodn}
{\bar\Psi}_{P'}(p)={\bar\Psi}_{P'}^0(p)+ig^2_{\pi
NN}\int\frac{d^4p'}{(2\pi)^4}\,
\Delta(p-p'){\tilde S}(p_2)\gamma_5{\bar\Psi}_{P'}(p')\gamma_5 S(p_1),
\end{equation}
where  $\Delta$ and  $S$ are the scalar and spinor
propagators, respectively,
${\tilde S}\equiv U_C\,S\,U_C^{-1}$, and  ${\bar\Psi}_{P'}^0(p)$ is the
free amplitude corresponding to two non interacting nucleons
(the relativistic plane wave). The solution of eq. (\ref{neodn})
may be presented as a Neuman-like series, the first term of which is the
free term from eq. (\ref{neodn}):
\begin{equation}
\label{rash}
{\bar\Psi}_{P'}(p)={\bar\Psi}_{P'}^0(p)+{\bar\Psi}_{P'}^i(p).
\end{equation}
The second part in eq. (\ref{rash}) is entirely determined by
the interaction and may be symbolically referred  to as scattered wave.
To determine the scattered wave in eq. (\ref{rash})
it is necessary to solve the BS equation of the type
(\ref{neodn}). Solving the BS equation in the continuum
is a much more  cumbersome procedure than, e.g. for the homogenous equation.
Besides difficulties encountered in solving the latter (singularities of
amplitudes, poles in propagators, cuts etc.) the former even does not allow
the Wick rotation  \cite{wick} to the Euclidian space,
and in the Minkowsky space there are no rigourous mathematical
methods of finding solutions \footnote{Actually there is one realistic
solution of the inhomogeneous BS equation in the ladder approximation,
obtained by Tjon~\cite{Tjonsol}.}. However, an approximate solution of
eq. (\ref{neodn}) may be obtained by applying the so-called
"one-iteration approximation" \cite{ourprc}. Within that one may obtain
a rather good estimate of the interaction term.

\section{The one-iteration approximation}

For a consistent
 relativistic analysis of the reaction (\ref{reaction})
 one should solve the BS equation for both bound state and scattering
 state with the same interaction kernel. We have found a numerical
 solution for the deuteron bound state with a realistic one-boson exchange
 potential \cite{solution}. The Bete-Salpeter equation, after a partial
 decomposition,
 has been solved numerically by using an iteration method. We found that the iteration
 procedure converges rather quickly if the trial function is properly chosen.
 In such case even after the first iteration the BS solution coincides with the exact
 one up to relative momentum $p\sim 0.6-0.7~GeV/c$. This circumstance becomes useful
 if one needs an approximate solution of the BS equation at not too large momenta
 $p \leq 0.5-0.7~ GeV/c$. This is just our case, since in reaction (\ref{reaction})
 the relative momentum of the $pp$-pair is expected to be rather small and
 the scattering part of the amplitude (\ref{rash}) can be obtained from
 the equation (\ref{neodn}) by one iteration.
 To solve eq. (\ref{neodn}) we proceed as follows (see also
 ref. \cite{ourprc}): i) for simplicity, in the inhomogeneous BS equation  we
 leave only the pseudo scalar isovector exchanges ($\pi$-mesons)
ii) by disregarding the dependence  upon $p_0$  in the meson propagator
 in eq. (\ref{neodn}) and then using the standard  representation
 of propagators via generalized Legandre polynomials $Q_l$,
 one obtains  for the main partial amplitude
\begin{eqnarray}
\nonumber
&&
\phi_\1s0(p_0,|\bp|)=\phi_\1s0^0(p_0,|\bp|)-\frac{g_{\pi NN}^2}{4\pi}
\frac{1}
{\left(\frac{\sqsf}{2}-E_p\right)^2-p_0^2}\times\\
&&\int\limits_0^\infty\frac{d|\bp'|}{2\pi}
\,\frac{|\bp'|}{|\bp|}\,\frac{1}{E_pE_{p'}}\left[(E_pE_{p'}-m^2)
Q_0({\tilde y}_\mu)-|\bp||\bp'|Q_1({\tilde y}_\mu)\right]
 u_{^1\!S_0}(s_f,|\bp'|),
\label{before_int}
\end{eqnarray}
where
${\tilde y}_\mu=\displaystyle\frac{\bp^2+\bp^{'2}+\mu^2}{2|\bp||\bp'|}$. In obtaining
(\ref{before_int}) the integration over $p_0'$ has been carried out in the
residium ${\tilde p}_0=\displaystyle\frac{\sqsf}{2}-E_{p'}$ and,
by definition, the BS wave function in the continuum is
\begin{eqnarray}
 u_{^1\!S_0}(s_f,|\bp'|)=\frac{g_\1s0({\tilde p}_0, |\bp'|)}{\sqsf-2E_{p'}},
 \label{u_contin}
\end{eqnarray}
where $g_\1s0$ is the vertex function (for more details see \cite{ourprc,nashi}).
Now if we restrict ourself to only one iteration in (\ref{before_int})
and take  the trial function (\ref{u_contin}) as a non relativistic solution
of the Schr\"odinger equation, e.g. the Paris wave function
$u^{NR}_{^1S_0}(s_f,|\bp'|)$, the BS amplitude is obtained by formula
\begin{eqnarray}
\phi_\1s0(p_0,|{\bf p}|)=\phi_\1s0^0(p_0,|{\bf p}|) -
\frac{G^{o.i.}(\tilde p_0,|{\bf p}|)}{\left(\frac{\sqsf}{2}-E_p\right)^2-p_0^2},
\label{foi}
\end{eqnarray}
where the "one-iteration"  BS vertex $G^{o.i.}(\tilde p_0,|{\bf p}|)$
is defined as
\begin{eqnarray}
&&
G^{o.i.}(\tilde p_0,|{\bf p}|)
=\frac{1}{\pi}\,\frac{g_{\pi NN}^2}{4\pi}
\left\{ \left[1-\frac{E_p}{m}\right]
\int\limits_0^\infty dr\, e^{-\mu r}j_0(pr)\,u^{NR}_{^1\!S_0}(r)
+\right.
\nonumber\\
&&
\left .\frac{|\bp|}{m E_p} \int\limits_0^\infty dr\,
\frac{u^{NR}_{^1\!S_0}(r)}{r}\,e^{-\mu r}\,(1-\mu r)\,j_1(pr)
\right\}.
\label{fin_oi}
\end{eqnarray}
Now from eqs. (\ref{foi}) and (\ref{fin_oi}) one may easily find the non relativistic
analogue of the obtained formulae. So, the free term in eq. (\ref{foi}) together
with the first term in eq. (\ref{fin_oi}) reflect the non relativistic equation
for the $^1S_0$ wave function, while the second term in (\ref{fin_oi}) turns out
to be corrections of  pure relativistic origin.

\section{Numerical results}
\begin{figure}[hb]  
\epsfxsize 5in
\centerline{\epsfbox{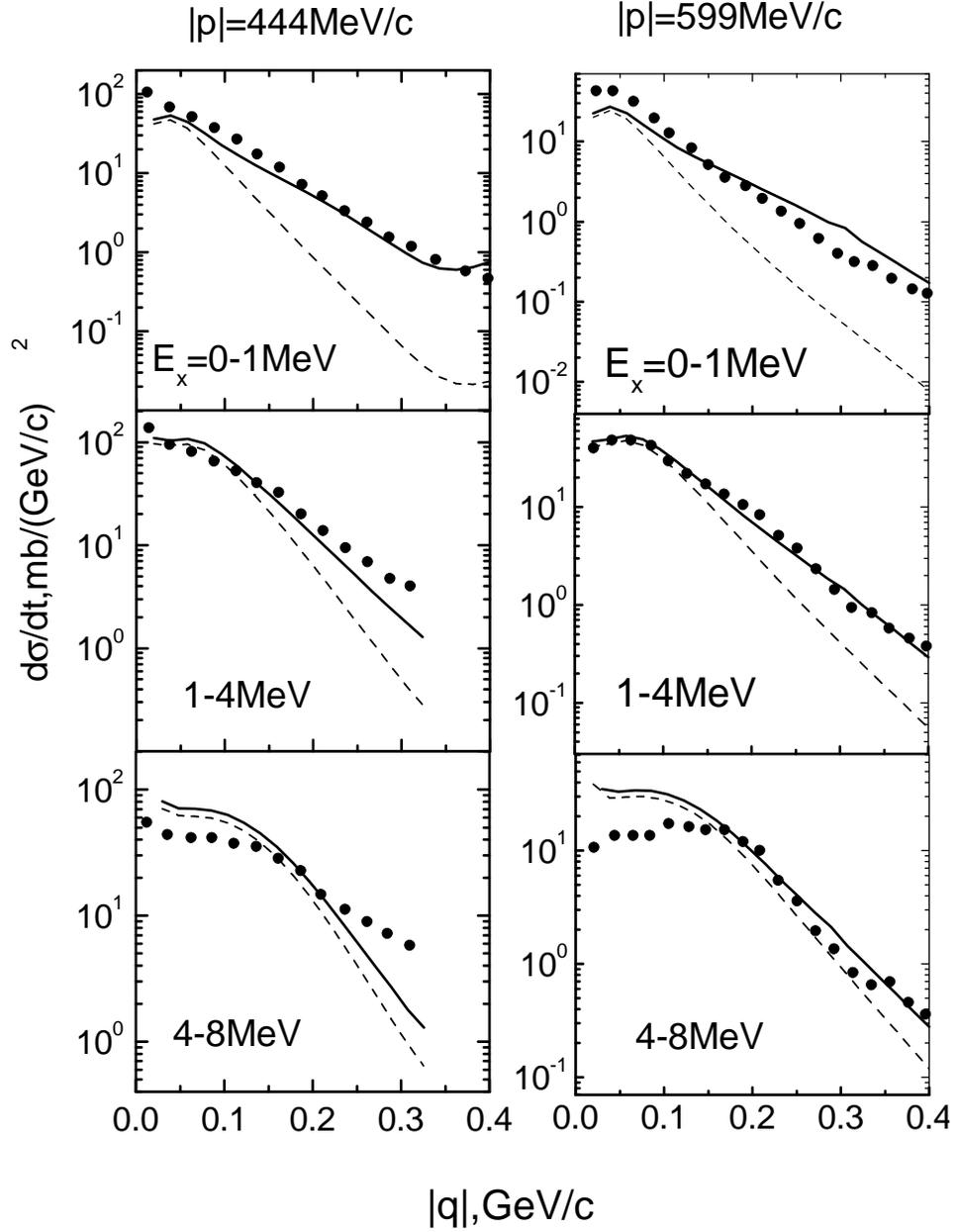}}
\vfill
\caption{ Results of full calculations of the differential cross section
 (\ref{crint}) with taking into account the effects of final
 state interaction in $^1S_0$ state (solid lines). Experimental
 data are those from SATURN-II~\cite{kox}, the elementary amplitude has been taken
 from  ref.\protect\cite{said,saidpap}. The dashed lines reflect
 the results of calculations within the  pure impulse approximation
 (cf. \cite{nash_yaf}).
}
\label{pict7}
\end{figure}

\begin{figure}[hb]  
\epsfxsize 5in
\centerline{\hspace*{-1cm} \epsfbox{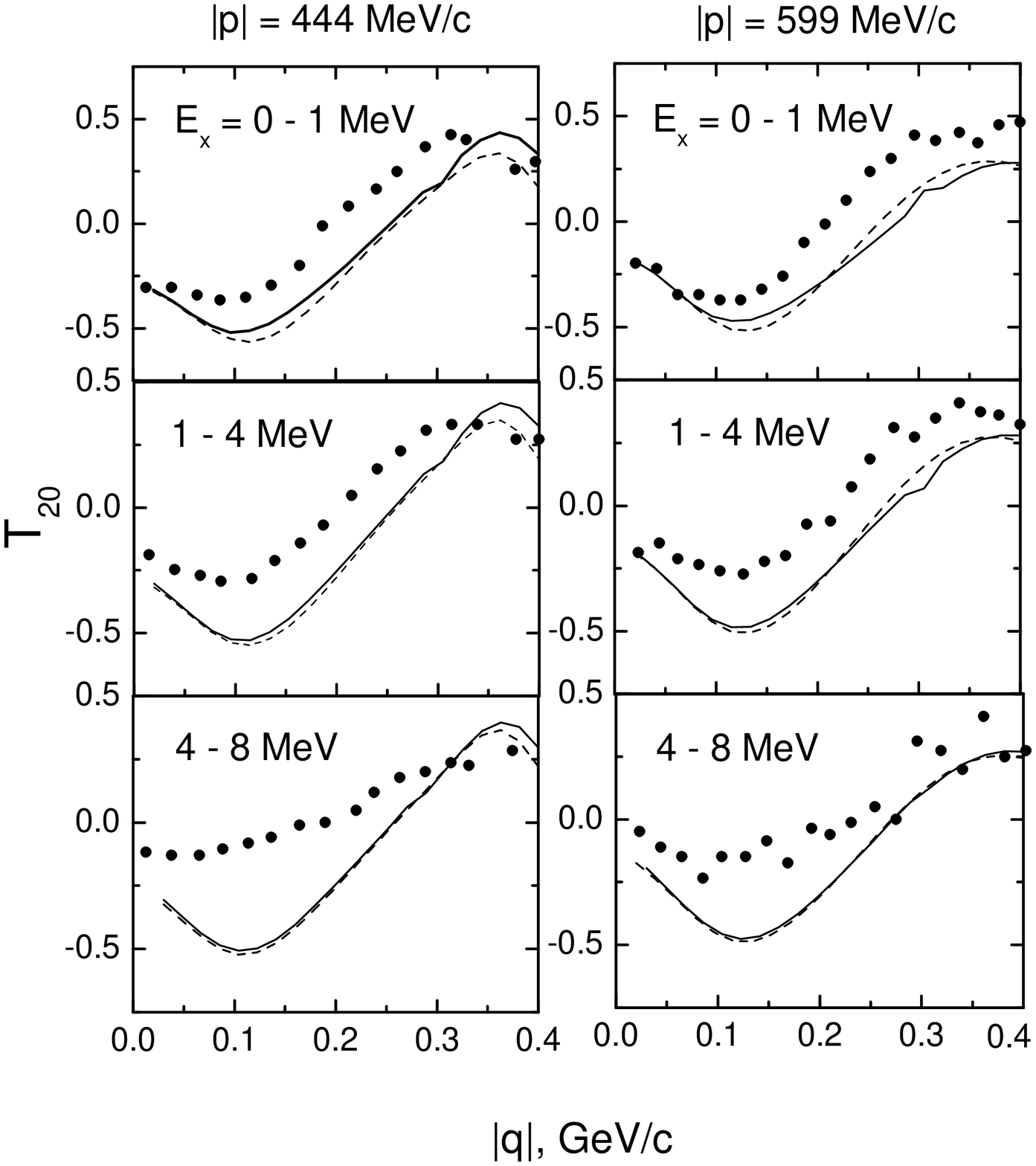}}
\vfill
\caption{Results of full calculations of the
 tensor analysing power
 with taking into account the effects of final
 state interaction in $^1S_0$ state. Notation as in Fig.~\ref{pict7}.
}
\label{pict8}
\end{figure}

\begin{figure}[hb]  
\centerline{\epsfbox{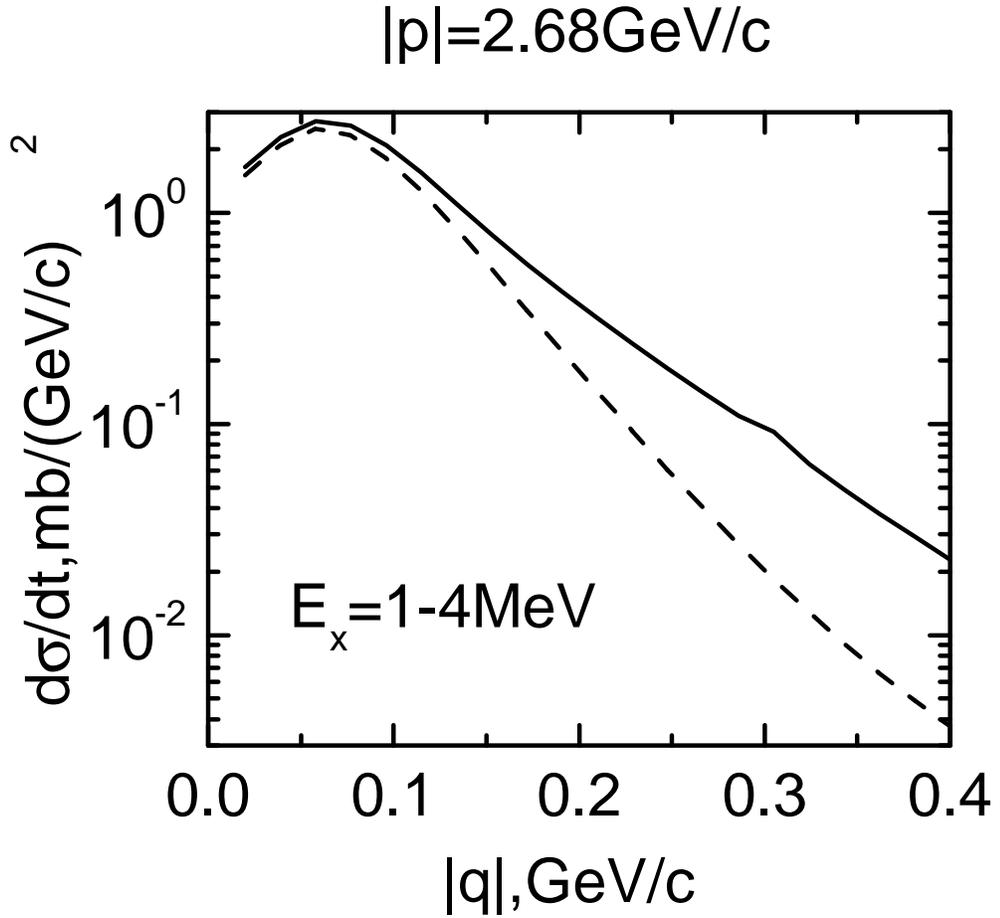}}
\vfill
\caption{ Results of full calculations of the differential cross section
 (\ref{crint}) with taking into account the effects of final
 state interaction in $^1S_0$ state (solid lines). Kinematical conditions
 correspond to those proposed at COSY~\cite{cosy_proposal}.
The elementary amplitude has been taken
 from  ref.\protect\cite{said,saidpap}.
 The dashed lines reflect  the results of calculations within the  pure impulse approximation
 (cf. \cite{nash_yaf}).
}
\label{pict9}
\end{figure}

\begin{figure}[hb]  
\centerline{\epsfbox{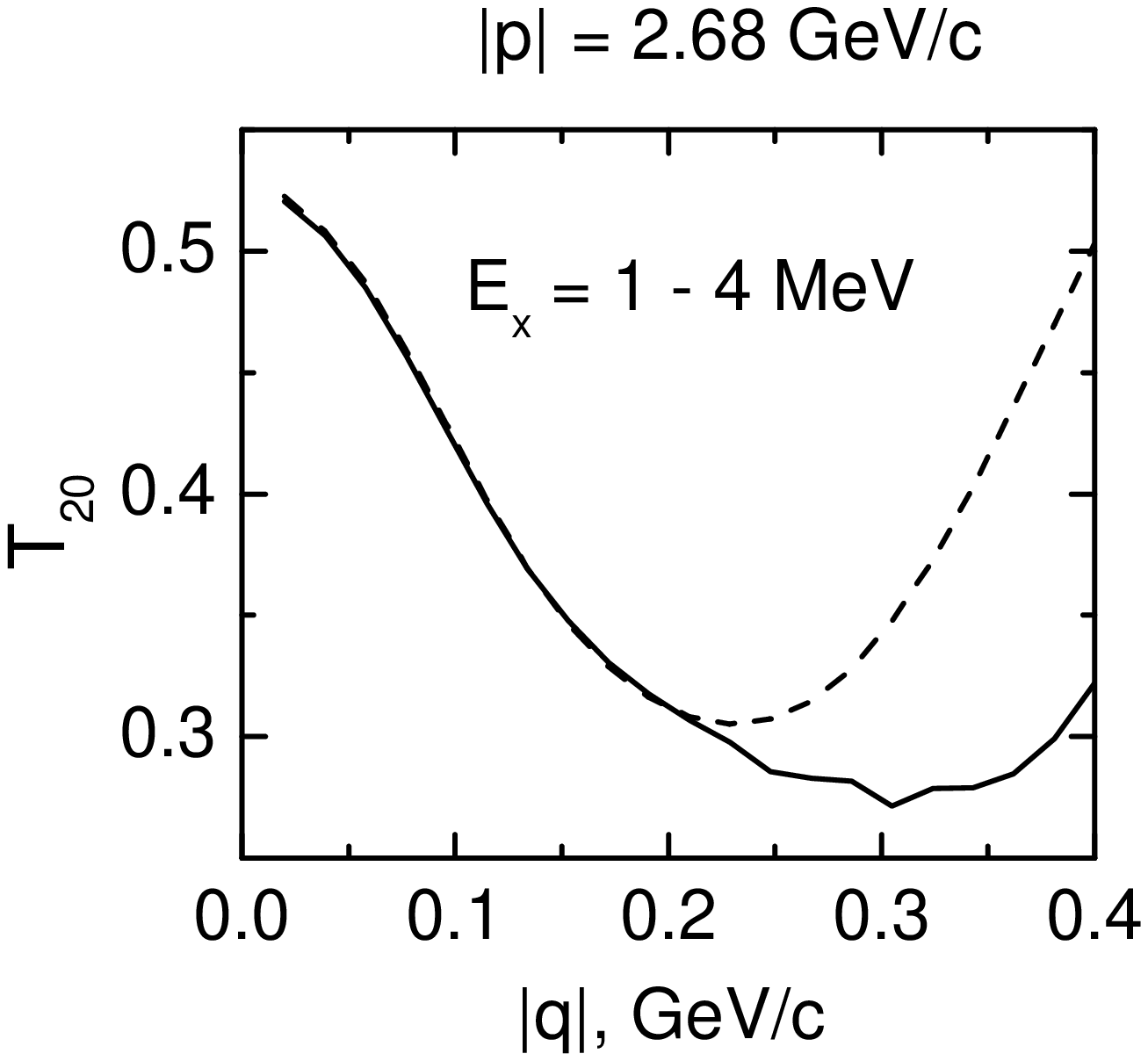}}
\vfill
\caption{ Results of full calculations of the
 tensor analysing power
 with taking into account the effects of final
 state interaction in $^1S_0$ state within the
 COSY kinematical conditions~\cite{cosy_proposal}.
 Other notation as in Fig.~\ref{pict9}.
}
 \label{pict10}
\end{figure}

In Figs.~\ref{pict7} and \ref{pict8} we present the
results of numerical calculations of
the cross section $\displaystyle\frac{d\sigma}{dt}$ and
tensor analysing power  $T_{20}$.
The elementary
charge exchange amplitude has been taken from ref.~\cite{saidpap} and
the non relativistic trial function $u^{NR}_{^1S_0}(r)$ is the
solution of the Schr\"odinger equation with the Paris potential~\cite{paris_cont}.
The Bethe-Salpeter amplitudes are those from
the  numerical solution \cite{solution}
 obtained with a realistic one-boson exchange interaction.
The dashed  lines in Figs. \ref{pict7} and \ref{pict8} correspond to
results within the relativistic impulse approximation \cite{nash_yaf},
while the solid lines
denote results with taking into account the final state interaction in
one-iteration approximation (as it is described above).
 It is seen that in all three energy bins the agreement with data for the cross section
 is essentially improved.  Especially it concerns the range $1\leq E_x\leq 4$. For
the  energy bin close to zero there is still a disagreement with data at low transferred
 momenta which, probably may be addressed to the fact that in our calculations
 we have not taken into account the Coulomb interaction in the $pp$-pair. For
  higher excitation energies ($E_x\sim 8 ~MeV$), other partial waves (e.g.
 triplet state ) in the $pp$-final state contribute and, within the adopted
 assumptions one may expect only qualitative agreement with data. From
 the Fig. \ref{pict7} one may conclude that at low excitation energies
 the supposed mechanism for the reaction (\ref{reaction}) (charge
 exchange subprocess with  interaction in $^1S_0$ state of the $pp$-pair
 in the continuum) seems to be correct. Moreover, from a comparison of
 the left and right panels in Fig.~\ref{pict7} one may expect that for the
 higher initial energy there is larger kinematical region where the mechanism
 holds. Figure \ref{pict8} demonstrates that the tensor analysing power
 is less sensitive to final state interaction effects. As a matter of fact,
 the tensor analysing power,
 being a ratio of non diagonal products of partial amplitudes
 to the  diagonal ones, serves as a measure of the quality of parametrization
 of partial amplitudes and their mutual relative phases.
 This has been pointed out
 in a series of publications (see e.g. refs.\cite{inclusive,ourprc}), where
 a good simultaneous description of
 cross sections and $T_{20}$ in reactions of the deuteron
 break-up or elastic scattering from protons, is still lucking.
 Nevertheless, since in the process (\ref{reaction}) the
 behaviour of the partial amplitudes
 is mostly governed  by the elementary
charge exchange ones, an experimental investigation
of the
tensor analysing power $T_{20}$ in reactions
of the type (\ref{reaction}) can essentially supplement  data on
the $NN$ charge exchange amplitudes at high energies.
In Figs.~\ref{pict9} and \ref{pict10} we present the predicted cross
section and tensor analysing power at high energies (COSY, Dubna).
It is immediately seen that the cross section is substantially
decreasing with the energy increasing, nevertheless remains large
enough to be experimentally investigated. Another peculiarity of the
studied process at relativistic energies is that the tensor analysing
power $T_{20}$ does not change the sign remaining positive in a large
kinematical region, in contrast to the lower energies (cf. Fig.
\ref{pict8}).
 Note again, that in the above calculations the vector analyzing
 power of the deuteron is strictly zero.

From the performed analysis one can conclude that there is a
kinematical region for the excitation energy,  $E_x <5~ MeV$, and
transferred momentum, $|{\bf q}| \le\, 0.3\div 0.4\,~ GeV/c$
(the COSY~\cite{cosy_proposal} kinematics), for which
the mechanism of the reaction (\ref{reaction}) is fairly well described
within the spectator approach by an elementary $pn$  charge exchange
subprocess, for active nucleons, with detection of the $pp$-pair in
the $^1S_0$ final state.
Our covariant approach agrees with previous non relativistic
calculations and allows predictions of the cross sections and
polarization observables at intermediate and relativistic energies,
in particular, for  kinematical conditions achievable at COSY.
The predicted cross sections and tensor analysing power
$T_{20}$ are large enough for the process (\ref{reaction}) to be
used, in a large range of initial energies,
for the determination of  properties of polarized deuteron, provided
experimentally one simultaneously detects vanishing  vector polarization
of deuterons.

\section{Summary} \label{gl4}

In summary, the performed covariant analysis of the
reactions  $\vec D(p,n)pp$ with two protons in the $^1S_0$ final state
allows to conclude that, as in the non relativistic limit, such process
can be used as an effective deuteron polarimeter also
at relativistic energies, achieved at COSY and Dubna.
The effects of final state interaction are found to be
substantial and essentially improve the agreement with data.

\section{Acknowledgments}
This work was performed in parts during the visits of S.S.S and L.P.K.
in the Forschungszentrum Rossendorf, Institute of Nuclear and Hadron Physics.
We thank for the support by the program "Heisenberg-Landau" of JINR-FRG
collaboration and the grants 06DR921, WTZ RUS 98/678 and RFBR 00-15-96737.


\newpage

\newpage

\newpage

\newpage

\newpage

\begin{thebibliography}{99}
\bibitem{alexa}
         L.C. Alexa, B.D. Adrson, K.A. Aniol et al., Phys. Rev. Lett.
         {\bf 82}, 1374 (1999).
\bibitem{kox0}
            S. Kox, E.J. Beise (spokespersons),
            TJNAF experiments 94-018 "Measurement of the Deuteron Polarization at
             Large Momentum Transvers in $D(e,e')D$ Scattering";
            Nucl. Phys. {\bf A684} 521, (2001).
\bibitem{preliminar}
            E. Tomasi-Gustafsson, in Proc. of {\it XIV International Seminar
            On High Energy Physics Problems}, Preprint JINR No. E1,2-2000-166
            (Dubna, 2000).
\bibitem{cosy_proposal}
           V.I. Komarov (spokesman) et al.,
        COSY proposal \#20 updated from 1999,
         ``Exclusive deuteron break-up study with
           polarized protons and deuterons at COSY''
           http://ikpd15.ikp.kfa-juelich.de:8085/doc/Proposals.html
\bibitem{elsatic}
           M.P. Rekalo, I.M. Sitnik, Phys. Lett. {\bf B 356}, 434 (1995);\\
           L.S. Azhgirei et al.,
           Phys. Lett. {\bf B 361}, 21 (1995).
\bibitem{inclusive}
          V.G. Ableev et al., Nucl. Phys. {\bf A 393}, 491 (1983);\\
          C.F. Perdrisat, V. Punjabi, Phys. Rev. {\bf C 42}, 1899 (1990);\\
          B. K\"uhn, C.F. Perdrisat, E.A. Strokovsky,
           Phys. Lett. {\bf B312}, 298 (1994).
\bibitem{cosy}
       A.K. Kacharava et al., Preprint JINR No. E1-96-42 (Dubna, 1996);
       C.F. Perdrisat (spokesperson) et al.,
           COSY proposal \#68.1 ''Proton-to-proton polarization transfer
          in backward elastic scattering''.
\bibitem{rekalo}
             M.P. Rekalo, N.M. Piskunov, I.M. Sitnik,
             Few Body Syst. {\bf 23}, 187 (1998).
\bibitem{ladygin}
           V.P. Ladygin, N.B. Ladygina, J. Phys. {\bf G 23}, 847 (1997).
\bibitem{ourprc}
      L.P. Kaptari, B. K\"ampfer, S.M. Dorkin, S.S. Semikh,
        Phys. Rev. {\bf C57}, 1097 (1998).
\bibitem{wil1}
          D.V. Bugg, C. Wilkin, Phys. Lett. {\bf B152}, 37 (1985);\\
          D.V. Bugg, C. Wilkin, Nucl. Phys. {\bf A467}, 575 (1987).
\bibitem{polder}
           S. Kox et. al., Nucl. Instrum. Methods {\bf A346}, 527 (1994).
\bibitem{ishida}
             Y. Satou, S. Ishida, H. Sakai, H.Okamura,  et al., Phys. Lett. {\bf B521},
             153, (2001).
\bibitem{moto}
           T. Motobayashi et al., Phys. Lett. {\bf B233}, 69 (1989).
\bibitem{wil2}
           J. Carbonell, M. Barbaro, C. Wilkin, Nucl. Phys. {\bf
           A529}, 653 (1991).
\bibitem{kox}
           S. Kox et. al., Nucl. Phys. {\bf A556}, 621 (1993).
\bibitem{bugg}
         D.V. Bugg, A. Hasan, R.L. Shypit, Nucl. Phys. {\bf
         A477}, 546 (1988);\\
         C. Furget et. al., Nucl. Phys. {\bf A631}, 747
         (1998).
\bibitem{morlet}
             M. Morlet et. al., Phys. Lett. {\bf B247}, 228 (1990).
\bibitem{nash_yaf}
            S.M. Dorkin, L.P. Kaptari, B. K\"ampfer, S.S. Semikh,
            Phys. Atom. Nucl. 65 (2002) 442; Yad. Fiz. 65 (2002) 469;
            see also nucl-th/0012088.
\bibitem{Tjonsol}
         J. Fleischer, J. Tjon, Nucl. Phys. {\bf B84}, 375 (1975);
         Phys. Rev. {\bf D15}, 2537 (1975)\\
         M. Zuilhof, J. Tjon, Phys. Rev. {\bf C22}, 2369 (1980).
\bibitem{solution}
         A.Yu. Umnikov, L.P. Kaptari, F.C Khanna,
         Phys. Rev. {\bf C56}, 1700 (1997); \\
         A.Yu. Umnikov, L.P. Kaptari, K.Yu.
         Kazakov, F.C. Khanna, Phys. Lett. {\bf B334}, 163 (1994).
\bibitem{parametrization}
         A.Yu. Umnikov, Z. Phys. {\bf A357}, 333 (1997).
\bibitem{mandel}
        S. Mandelstam, Proc. Roy. Soc. (London) {\bf A233}, 123 (1955).
\bibitem{nashi}
        S.G. Bondarenko, V.V. Burov, M. Beyer, S.M. Dorkin,
        Phys. Rev. {\bf C58}, 3143 (1998).
\bibitem{wick}
        G.C. Wick, Phys. Rev. {\bf 96}, 1124 (1954).
\bibitem{www}
        http://nn-online.sci.kun.nl
\bibitem{said}
        http://said.phys.vt.edu
\bibitem{swart}
        V.G.J. Stoks, R.A.M. Klomp, M.C.M. Rentmeester, J.J. de Swart,
        Phys. Rev. {\bf C48}, 792 (1993).
\bibitem{saidpap}
        R.A. Arndt, I.I. Strakovsky, R.L. Workman,
        Phys. Rev. {\bf C62}:034005, 2000
\bibitem{paris_cont}
        M. Lacombe, B. Loiseau, J.M. Richard, R. Vinh Mau et al., Phys. Rev. {\bf C21},
       861 (1980)
\end{thebibliography}
\end{document}